# DISSIPATIVE PROPERTIES OF ISOSCALAR COLLECTIVE EXCITATIONS IN HEAVY NUCLEI

**V. I. Abrosimov[*] and O. I. Davydovska**

*Institute for Nuclear Research, National Academy of Sciences of Ukraine, Kyiv, Ukraine*

[*]Corresponding author: abrosim@kinr.kiev.ua

## Abstract

The dissipative properties of isoscalar collective motion in heavy nuclei are studied within a kinetic model of collective excitations in finite Fermi systems with a free moving surface. The model includes both the one-body damping produced by the interaction of nucleons with the dynamic surface of nucleus and the Landau damping mechanism. In this paper, we study the damping and dissipative properties of the isoscalar dipole collective modes and the isoscalar vortex octupole mode, which are formed due to dynamic effects of the nuclear surface. To study the damping mechanism of the considered collective modes, a multipole strength function is used in various approximations. It is shown that the damping of the considered isoscalar collective modes is due to a one-body damping mechanism. An analytical expression for the dissipation coefficient at finite frequency for the collective variable describing multipole oscillations of nuclear surface is obtained. It was found that the dissipation coefficient in the energy region ($E=\hbar\omega$) of the considered isoscalar collective modes does not exceed the wall formula friction coefficient derived in the one-body dissipation theory. This result confirms the understanding of friction in the wall formula as a macroscopic limit of the one-body nuclear dissipation.

## 1. Introduction

It is known that isoscalar collective excitations in nuclei at excitation energies of the order of 10 MeV and more are damped and exhibit dissipative properties [1], but the damping mechanism and the character of dissipation depend on the nature of the collective excitation. In previous papers [2, 3], we studied isoscalar dipole and octupole collective excitations in heavy spherical nuclei in a kinetic model of collective excitations in finite Fermi systems with a free moving surface. It was found that the dynamic effects of the nuclear surface have an essential influence on the formation of these isoscalar collective excitations in nuclei.

In this paper, we study the damping and dissipative properties of the isoscalar dipole collective modes and the isoscalar vortex octupole mode, which are formed due to dynamic effects of the nuclear surface. We will focus on the collisionless damping mechanism of the isoscalar collective excitations under consideration and use the kinetic model of isoscalar collective excitations in finite Fermi systems with a free moving surface [4]. This self-consistent model contains both the one-body damping arising due to the interaction of nucleons with the dynamic surface of nucleus and the Landau damping mechanism. The kinetic model allows us to consider the collisionless damping of collective excitations in different approximations. The fixed-surface model does not contain the one-body damping mechanism, while the moving-surface model, without taking into account the residual interaction between nucleons, does not have the Landau damping.

In Section 2, we discuss the mechanisms of collisionless damping of isoscalar collective excitations contained in our kinetic model. In order to study the damping of the isoscalar collective modes under consideration, we use the multipole strength function in different approximations. In Section 3, to consider the dissipative properties of isoscalar collective modes, we derive the dissipation coefficient at finite frequency for collective variable that describes the axially-symmetric multipole oscillations of nuclear surface in our model.

## 2. Strength function and one-body damping of collective excitations

We will use the kinetic model of isoscalar collective excitations in nuclei [4]. In this kinetic model, the nucleus is considered as a gas of interacting nucleons confined to a spherical cavity with a free moving surface. The main dynamical quantity of the model is the variation of the phase-space distribution function $\delta n(\boldsymbol{r},\boldsymbol{p},t)$ related to the collective excitation. This quantity is determined by the (linearized) Vlasov equation

$$\frac{\partial}{\partial t}\delta n(\boldsymbol{r},\boldsymbol{p},t)+\frac{\boldsymbol{p}}{m}\frac{\partial}{\partial \boldsymbol{r}}\left[\delta n(\boldsymbol{r},\boldsymbol{p},t)-\frac{dn_0}{d\varepsilon}(\delta V(\boldsymbol{r},t)+V_{ext}(\boldsymbol{r},t))\right]=0, \tag{1}$$

the equation of surface motion $R(\theta,\varphi,t)$, which is a sphere with radius $R=r_0A^{1/3}$ in equilibrium, and whose oscillations are described by collective variables $\delta R_{\mathrm{LM}}(t)$,

$$R(\vartheta,\varphi,t)=R+\sum_{LM}\delta R_{LM}(t)\,Y_{LM}(\vartheta,\varphi) \tag{2}$$

and the boundary conditions at the moving surface that ensure a free dynamic surface: the mirror reflection condition of particles from the moving surface and the boundary condition that provides the consistent motion of particles inside the nucleus and the dynamic surface (the self-consistency boundary condition). It should be noted that the right-hand side of the kinetic equation (1) is zero and therefore the model does not take into account the effects associated with collisions between nucleons. The kinetic equation (1) contains the derivative of the equilibrium density distribution function with respect to the single-particle energy $dn_0/d\varepsilon$. In our kinetic model, we assume that the function $n_0$ is a step function of the single-particle energy ε, which terminates at the Fermi energy $\varepsilon_F$. In the kinetic equation (1), the change of the mean field $\delta V(\boldsymbol{r},t)$ is generated by the residual interaction between nucleons $v(\boldsymbol{r},\boldsymbol{r}')$ and is determined as follows

$$\delta V(\boldsymbol{r},t) = \int d\boldsymbol{r}'\, d\boldsymbol{p}\; v_L(\boldsymbol{r},\boldsymbol{r}')\,\delta n(\boldsymbol{r}',\boldsymbol{p},t)\,. \tag{3}$$

In order to obtain an explicit (analytic) solution of the dynamic equations (1)-(2), we will assume a separable multipole residual interaction of the form

$$v_L(\boldsymbol{r},\boldsymbol{r}') = \kappa_L \sum_M r^{L+k} r'^{L+k} Y_{LM}(\vartheta,\varphi)\, Y^*_{LM}(\vartheta',\varphi')\,, \tag{4}$$

where $\kappa_L$ is a parameter that determines the strength of the isoscalar multipole interaction, and k=2 for L=0.1 and k=0 for L≥2.

The solution of the linearized Vlasov equation (1) for the variation of the phase-space distribution function can be written as the sum:

$$\delta n(\boldsymbol{r},\boldsymbol{p},t) = \delta n_{stat.surf}(\boldsymbol{r},\boldsymbol{p},t) + \delta n_{dyn.surf}(\boldsymbol{r},\boldsymbol{p},t)\,. \tag{5}$$

Here, the first component $\delta n_{stat.surf}(\boldsymbol{r},\boldsymbol{p},t)$ is a solution for a system bounded by a static surface, when the right-hand side of the equation of motion of nuclear surface (2) is equal to the equilibrium radius $R$ ($\delta R_{LM}(t)=0$), while the second component is proportional to the amplitude of surface oscillations $\delta n_{dyn.surf}(\boldsymbol{r},\boldsymbol{p},t) \sim \delta R(\upsilon,\varphi,t)$ and is due only to the dynamic surface effects.

The model includes a mechanism of the collisionless Landau damping, which has the same origin as the collisionless Landau damping of zero sound in an infinite Fermi system [5]. The Landau damping is related to the component $\delta n_{stat.surf}(\boldsymbol{r},\boldsymbol{p},t)$ in equation (5) and arises when the residual interaction between nucleons is taking into account: $\delta V(\boldsymbol{r},t) \neq 0$ in the kinetic equation (1). In addition, the model also includes a one-body damping mechanism arising from the interaction of nucleons with the dynamic surface of nucleus. Indeed, the study of the surface response of a semi-infinite Fermi system within the kinetic model under consideration revealed a damping surface mode, the friction coefficient of which coincides with friction in the wall formula [6], so it can be assumed that the damping effects due to the one-body friction are also present in the kinetic model for a finite Fermi system. In our self-consistent kinetic model, one would expect that there is a connection between the collisionless damping mechanisms under consideration.

The solution of the dynamic equations (1)-(2) of the kinetic model for isoscalar multipole collective excitations in spherical nuclei can be found using the linear response function formalism. For this, we assume that isoscalar axially-symmetric multipole excitations are excited by a weak external field, which has the form

$$V_{ext}(\boldsymbol{r},t) = \beta_L\, \delta(t)\, r^{L+k}\, Y_{L0}(\vartheta,\varphi) \tag{6}$$

where $\beta_L$ is a small parameter describing the strength of the external field, $\delta(t)$ is the Dirac delta-function in time, k=2 for L=0.1 and k=0 for L≥2.

By using the solution, we can find an isoscalar multipole response function, the imaginary part of which (the strength function) determines the isoscalar collective excitations, in particular the centroid energies and the widths of the collective excitations. Our model allows us to study collective excitations in different approximations, in particular in fixed- and moving-surface models. We consider the isoscalar multipole strength function defined as

$$S_L(E) = -\frac{1}{\pi} Im R_L(E)\,. \tag{7}$$

Here $E=\hbar\omega$ is the excitation energy, and the isoscalar multipole response function $R_L(E)$ is defined as

$$R_L(\omega) = \beta_L^{-1} \int d\boldsymbol{r}\, r^{L+k}\, Y_{L0}^*(\vartheta,\varphi)\, \delta\bar{\varrho}_L(\boldsymbol{r},\omega) \qquad , \tag{8}$$

where the particle density variation $\delta\bar{\varrho}_L(\mathbf{r},\omega)$ is determined taking into account the moving surface and therefore it has an additional term

$$\delta\bar{\varrho}_L(\boldsymbol{r},\omega) = \delta\varrho_L(\boldsymbol{r},\omega) + \varrho_0\, \delta(r-R)\, \delta R_{L0}(\vartheta,\varphi,\omega)\,. \tag{9}$$

In equation (9) $\rho_0$ is the equilibrium nuclear density, and

$$\delta\varrho_L(\boldsymbol{r},\omega) = \int d\boldsymbol{p}\ \delta n_L(\boldsymbol{r},\boldsymbol{p},\omega)\,, \tag{10}$$

where $\delta n_L$ is the variation of the phase-space distribution function related to isoscalar multipole excitations, which is obtained in our kinetic model. To study the damping of the isoscalar collective modes under consideration, we will use the multipole strength function (7) in different approximations.

Numerical calculations of isoscalar dipole and octupole strength functions in different approximations were carried out. Fig. 1 shows the isoscalar dipole strength function in different approximations. Fig. 1(a) shows the dipole strength function in models with a moving (solid curve) and a fixed surface (dashed curve) without taking into account the residual interaction (the strength parameter $\kappa$ is zero), while Fig. 1(b) presents the strength function in the model with a moving surface taking into account the interaction between nucleons, the strength parameter $\kappa \neq 0$ (dashed curve), and also at $\kappa = 0$ (solid curve). It can be seen that the resonance structures around the energy of 12 MeV (vortex dipole mode) and 25 MeV (high-energy dipole mode) are formed in the moving surface model without taking into account the residual interaction between nucleons (solid curves), so their damping (width) is due to the one-body damping mechanism. When the residual interaction between nucleons is taken into account, the centroids of these resonances are slightly shifted, and the widths of the resonances are reduced (compare the solid and dashed curves). The following values of nuclear parameters were used in the calculations: $r_0$=1.25 fm, $\varepsilon_F$ = 30.94 MeV and m =1.04 MeV $(10^{-22}\text{sec})^2/\text{fm}^2$. The parameter of the isoscalar dipole interaction strength, see equation (4) at L=1, is chosen to be equal to MeV/fm$^2$ in order to reproduce the experimental value of the giant monopole resonance energy in the $^{208}$Pb nucleus within the framework of our kinetic model [3].

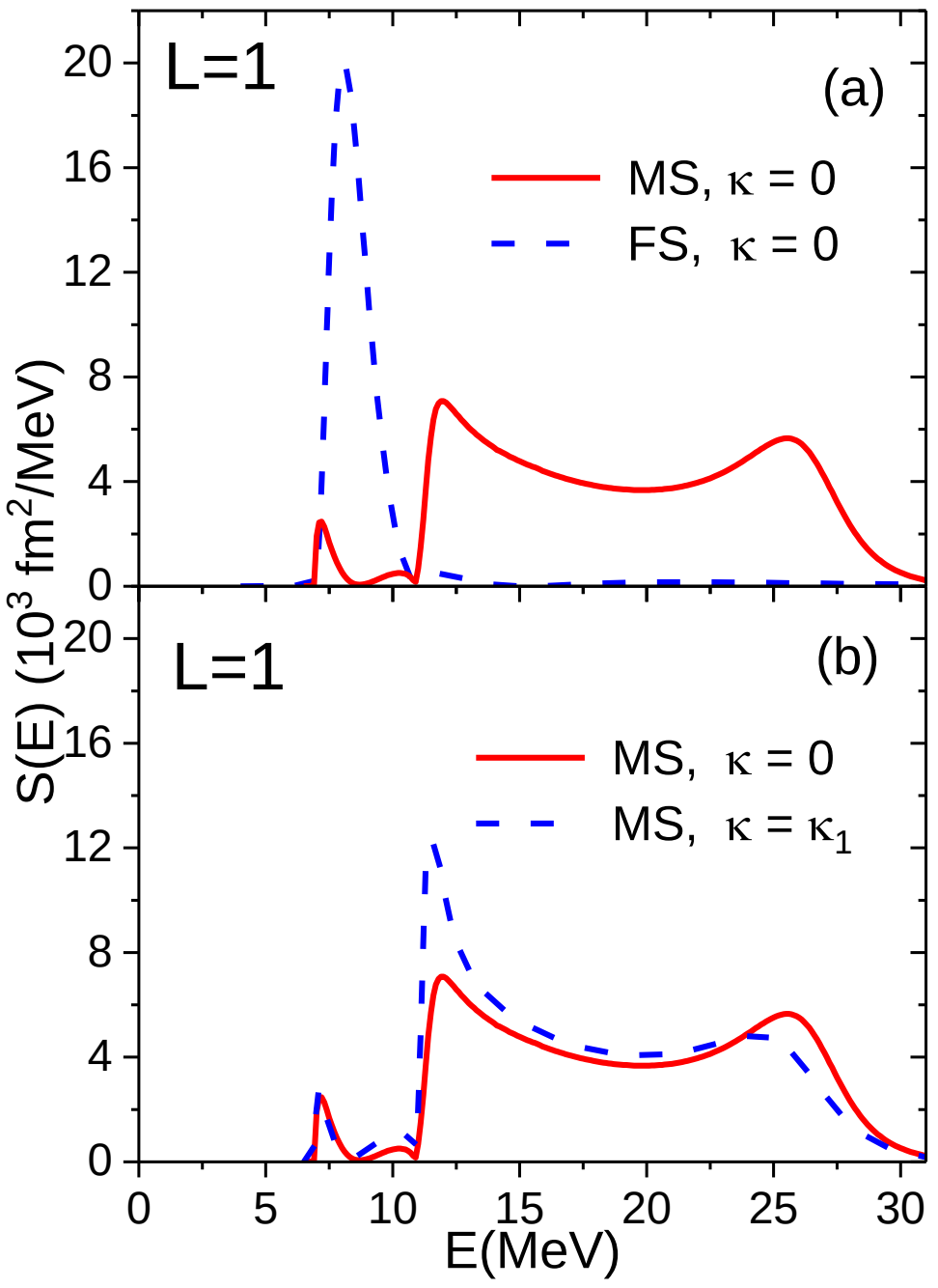

Fig. 1. Isoscalar dipole strength function (L=1) in different approximations. In Fig. 1(a), the strength function is shown in moving- (solid curve) and fixed-surface models (dashed curve), the strength parameter $\kappa = 0$. In Fig. 1(b), the moving-surface dipole strength function is given for the strength parameters: $\kappa = 0$ (solid curve), and $\kappa = \kappa_L$ (dashed curve). The parameter $\kappa_L$ is chosen as $\kappa_1 = -7.5 \cdot 10^{-3}$ MeV/fm$^2$. The system contains A=208 nucleons.

The isoscalar octupole strength function is shown in Fig. 2. The isoscalar octupole strength function exhibits similar properties for the resonance structure around 12.5 MeV (vortex octupole mode). It can be seen that this resonance structure is formed in the moving surface model without taking into account the residual interaction between nucleons (solid curves), therefore, its width is due to the one-body damping mechanism. When the residual interaction between nucleons is taken into account (the strength parameter $\kappa \neq 0$), the centroid of the resonance shifts slightly and the resonance width decreases (compare the solid and dashed curves in Fig. 1(b)). It should be noted that the residual interaction dramatically changes the response in the high-energy octupole resonance region (around 20.2 MeV). The following values of nuclear parameters were used in the calculations: $r_0$=1.2 fm, $\varepsilon_F$ = 33.42 MeV and m =1.04 MeV $(10^{-22}\text{sec})^2/\text{fm}^2$. The parameter of the isoscalar octupole interaction strength, see equation (4) at L=3, is chosen to be $\kappa_3$=-2.10$^{-5}$ MeV/fm$^6$ to reproduce the experimental value of the centroid energy of the high-energy octupole resonance in the $^{208}$Pb nucleus [2].

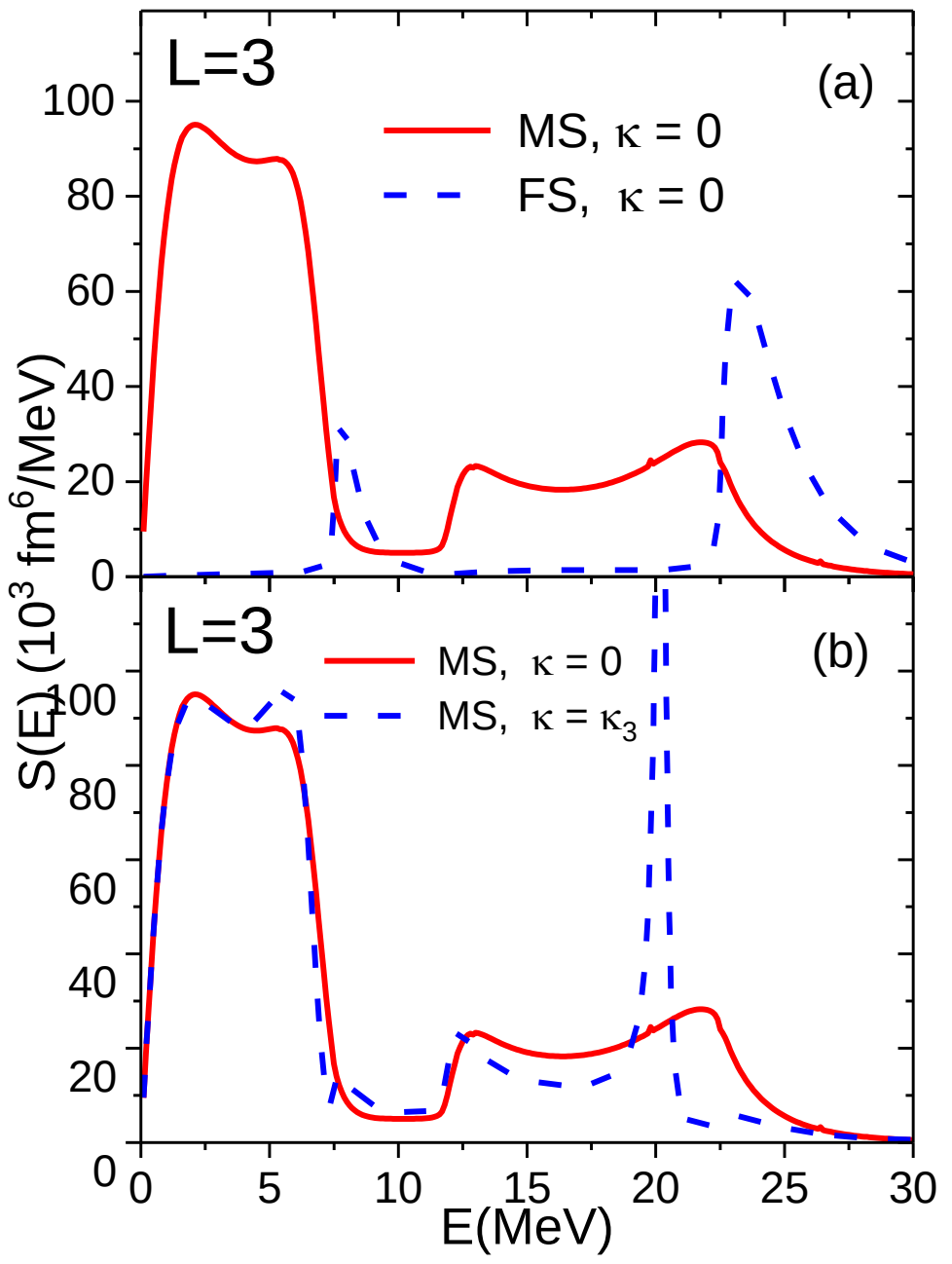

Fig. 2. Isoscalar octrupole strength function (L=3) in different approximations. In Fig. 1(a), the strength function is shown in moving- (solid curve) and fixed-surface models (dashed curve), the strength parameter $\kappa = 0$. In Fig. 1(b), the moving-surface strength function is given for the strength parameters: $\kappa = 0$ (solid curve), and $\kappa = \kappa_L$ (dashed curve). The parameter $\kappa_L$ is chosen as $\kappa_3 = -2 \cdot 10^{-5}$ MeV/fm$^6$. The system contains A=208 nucleons.

The results shown in Fig. 1 and 2 demonstrate that the collisionless damping of the isoscalar dipole collective modes and the isoscalar vortex octupole mode, which are formed due to dynamic effects of the nuclear surface, has a similar nature.

### 3. Dissipation coefficient at a finite frequency

In order to study the dissipative properties of isoscalar multipole collective excitations in nuclei, we consider the rate of energy dissipation that nucleus absorbs from a weak periodic external field $F_{ext}$ with the frequency $\omega$, which acts on the nuclear surface. The expression for this quantity can be written as follows:

$$\dot{Q}_L = \gamma_L(\omega) u_L^2(t) \quad , \tag{11}$$

where

$$u_L(t) = \frac{d}{dt} \delta R_{L0}(t) / R \tag{12}$$

is the normal component of the surface velocity for the collective variable $\delta R_{L0}(t)$, which describes the axially-symmetric multipole surface vibrations of spherical nucleus of radius $R$. In equation (11), $\gamma_L(\omega)$ is the dissipation coefficient for the collective variable $\delta R_{L0}(t)$, which characterizes the intensity of the dissipation process. We want to determine this dissipation coefficient by using our kinetic model.

We will use the linear response function formalism and therefore all physical quantities are determined by the corresponding response functions, in particular, the solution for the collective variable $\delta R_{L0}(t)$, which is a real quantity, is determined by the multipole surface response function $R_{L,surf}(\omega)$ as

$$\delta R_L(t) = \mathrm{Re}[\delta R_L(\omega)\exp(i\omega t)], \tag{13}$$

where

$$\delta R_L(\omega)/R = R_{L,surf}(\omega) F_{ext}(\omega). \tag{14}$$

In order to obtain the multipole surface response function within our kinetic approach, we assume that the isoscalar multipole modes are excited by a weak periodic external field with the frequency $\omega$, which acts on the nuclear surface. In this approximation, the external field does not come in the kinetic equation (1), but acts on the surface through the self-consistency boundary condition [4], which has the following form

$$\delta\,\Pi_{rr}(\boldsymbol{r},t)\,|_{\,r=R} = \delta P_\sigma(\vartheta,\varphi,t) - P_{ext}(\vartheta,\varphi,t), \tag{15}$$

The boundary condition (15) provides a free dynamic surface and determines the collective motion in nucleus. In equation (15), the normal component of the momentum flux tensor $\delta\,\Pi_{rr}(\mathbf{r},t)$, associated with radial forces inside of nucleus, describes the forces acting on the nuclear surface. The pressure $\delta P_\sigma(\vartheta,\varphi,t)$, due to surface tension forces arising from surface deformation, has the form

$$\delta P_\sigma(\vartheta,\varphi,t) = C_L\,\frac{\delta R_{L0}(t)}{R^4}\,Y_{L0}(\vartheta,\varphi), \tag{16}$$

where $C_L = \sigma(L-1)(L+2)R^2$ and $\sigma$ is the surface tension parameter (for the nuclear surface $\sigma \approx 1$ MeV/fm$^2$ [6]). The external pressure in equation (15) is chosen as:

$$P_{ext}(\vartheta,\varphi,t) = \frac{F_{ext}(t,\omega)}{R^3}\,Y_{L0}(\vartheta,\varphi). \tag{17}$$

Here, the function $F_{ext}$ (t, ω) describes a periodic external field with frequency $\omega$ acting on the nuclear surface. Using (15), we can find the multipole surface response function, which is given by equation (14), in the form

$$R_{L,surf}(\omega) = [C_L - \chi_L(\omega)]^{-1}. \tag{18}$$

In equation (18), the function $\chi_L(\omega)$ (see eq. (80) in [3]) determines how a gas of nucleons confined to a spherical cavity responds to the periodic multipole vibrations of the surface with the frequency $\omega$. The response function was found without taking into account the residual interaction between nucleons, since the residual interaction has an slight effect on the formation of the isoscalar collective excitations under consideration.

Further, using the surface response function (18), we can obtain the rate of energy transfer to nucleus from the external field $F_{ext}$, averaged over the period of vibrations of the external field T=2π/ω. This quantity is determined by the imaginary part of the response function (18) and has a standard form in models that use the linear response function formalism, see, e.g., [7], namely

$$\langle dE_L / dt \rangle_T = (1/2) F_{ext}^2(\omega)\omega \, R''_{L,surf}(\omega) . \tag{19}$$

Equating the rate of energy dissipation (11), averaged over the period of oscillation of the external field $F_{ext}$ (t, ω), to the average rate of energy transfer (19)

$$\langle \dot{Q}_L \rangle_T \ = \ \langle dE_L / dt \rangle_T \tag{20}$$

and taking into account equations (11) - (14) and (18), (19), we obtain the dissipation coefficient at finite frequency $\gamma_L(\omega)$ in the form

$$\gamma_L(\omega) = \frac{1}{\omega} Im \chi_L(\omega) . \tag{21}$$

The dissipation coefficient (21) is determined by the imaginary part of the function $\chi_L(\omega)$ . In the theory of the one-body dissipation [8, 9], the friction coefficient in the wall formula characterizes the rate of energy dissipation in a gas of nucleons confined in a cavity with a slowly moving surface, assuming complete randomization (randomness) of the nucleon motion. This coefficient for a slow, small-amplitude surface motion of a spherical system of radius $R$ has the form [8, 10]: $\gamma_{wf} = 3/4 \rho_0 p_F R^4$, where $\rho_0$ is the nuclear equilibrium density, and $p_F$ is the Fermi momentum. It can be seen that the friction coefficient in the wall formula does not depend on the multipolarity of the surface oscillations. The dissipation coefficient (21) can be written in terms of the friction coefficient in the wall formula as

$$\gamma_L(\omega) = \gamma_{wf} \ \Phi_L(\omega) . \tag{22}$$

In equation (22), the dimensionless function $\Phi_L(\omega)$ determines the dependence of the dissipation coefficient on the energy ($E=\hbar\omega$) of isoscalar multipole collective excitations, as well as on the multipolarity of the excitation $L$. It should be noted that at zero frequency the dissipation coefficient (21) coincides with the friction coefficient for low-energy multipole isoscalar collective excitations in nuclei, which was obtained in [11] in an analytical form (see equation (4.15) in [11]). It can be found that for the dipole modes $\gamma_{L=1}(\omega = 0) / \gamma_{wf} = 0$, while for the octupole ones $\gamma_{L=3}(\omega = 0) / \gamma_{wf} = 0{,}85$.

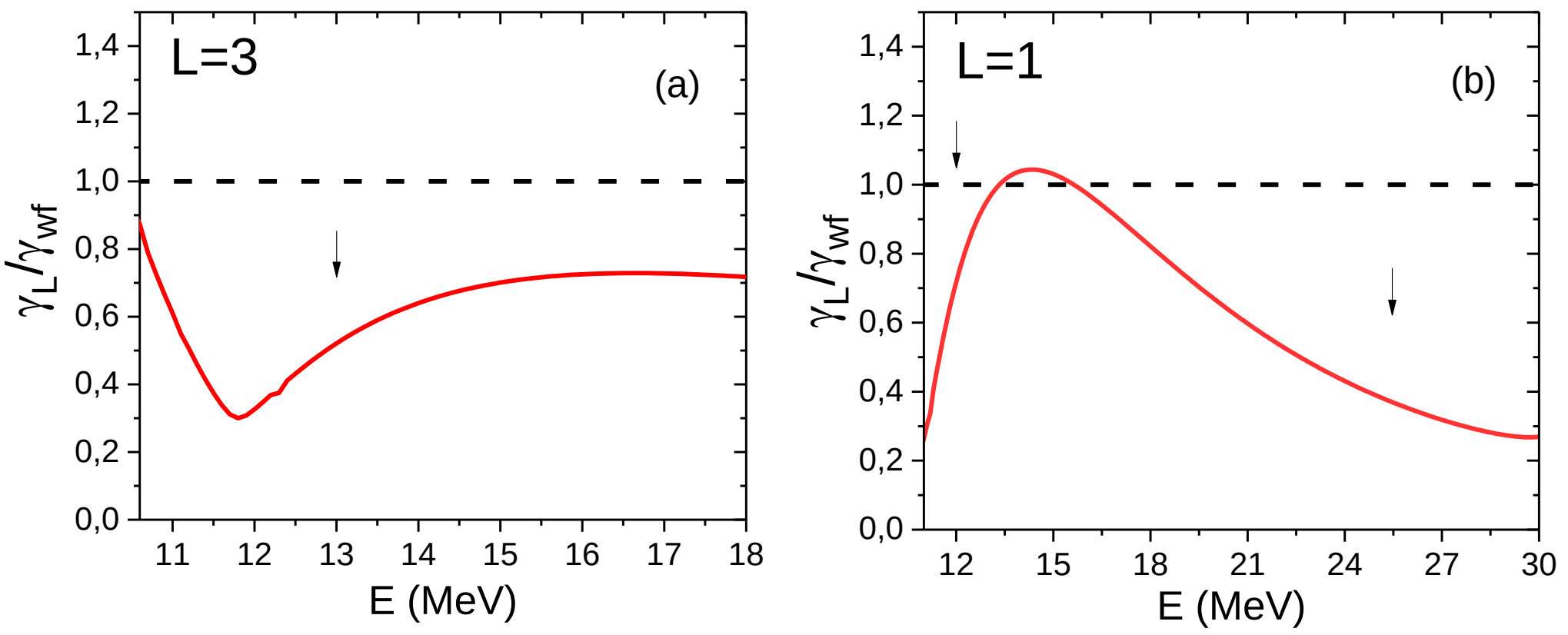

Fig. 3. The octupole and dipole dissipation coefficients $\gamma_L(\omega)$ are shown in units of the wall formula friction coefficient $\gamma_{wf}$ as a function of the excitation energy $E = \hbar\omega$. Fig. 3(a) shows the octupole dissipation coefficient in the energy region of the isoscalar vortex octupole mode, while Fig. 3(b) is for the dipole dissipation coefficient in the energy region of isoscalar dipole collective modes. The theoretical centroid energies of the isoscalar vortex octupole mode and the low- and high-energy isoscalar dipole collective modes (see Figs. 1, 2) are indicated by arrows.

Next, we consider the results of the numerical calculation of the dissipation coefficient (21) for the isoscalar dipole and octupole collective excitations. Fig. 3 shows the octupole and dipole dissipation coefficients in units of the friction coefficient in the wall formula in the energy range of the considered isoscalar collective modes (solid curves). The dashed line indicates the friction coefficient in the wall formula. Fig. 3(a) shows the octupole dissipation coefficient in the energy region of the isoscalar vortex octupole mode, while Fig. 3(b) shows the dipole dissipation coefficient in the energy region of the isoscalar dipole collective modes. The theoretical energies of the centroids of the octupole mode and the low- and high-energy dipole modes (see Figs. 1 and 2) are indicated by arrows. It can be seen from the figures that the dissipation coefficients for the vortex octupole mode with the centroid energy of 13 MeV, see Fig. 3(a), and the vortex dipole mode with the centroid energy of 12 MeV, see Fig. 3(b), exhaust about 61% and 72% of the friction coefficient in the wall formula, respectively, while the dissipation coefficient for the high-energy dipole mode with the centroid energy of 25.5 MeV, see Fig. 3(b), exhausts 37% of the friction coefficient in the wall formula. These results confirm our understanding of the friction coefficient in the wall formula as a macroscopic limit of the one-body nuclear dissipation.

## 4. Conclusions

We have studied the collisionless damping and dissipative properties of collective excitations in nuclei, namely the isoscalar dipole collective modes and the vortex octupole mode, caused by dynamic effects of the nuclear surface. For this, a self-consistent kinetic model of collective excitations in finite Fermi systems with a free moving surface was used.

This paper shows that the main contribution to the collisionless width of isoscalar collective excitations under consideration is due to the one-body damping mechanism. It was found that in a self-consistent model, the coupling between the one-body damping mechanism and the Landau one leads to a decrease in the width of the vortex octupole mode and isoscalar dipole modes in heavy nuclei.

We have evaluated the one-body dissipation coefficient in the energy region of isoscalar collective excitations under consideration. It was found that its value for both the isoscalar vortex octupole mode and the isoscalar dipole modes, which are formed due to dynamic effects of the nuclear surface, does not exceed the wall formula friction coefficient. This result confirms our understanding of the wall formula friction coefficient as a macroscopic limit of the one-body nuclear dissipation.

For the further investigation of the collisionless damping of collective excitations in nuclei, it is of interest to study the collisionless damping of isoscalar collective excitations of a different nature than those considered in this paper, in particular giant multipole resonances formed by the residual interaction between nucleons.